\documentclass[12pt,authoryear]{elsarticle}
\usepackage[top=1.0in,right=1.0in,left=1.0in,bottom=1.75in]{geometry}

\usepackage{color}

\usepackage{amssymb}
\usepackage{fancyhdr}            % NB: E=pagina pari, O=pagina dispari, H=intestazione, F=piede pagina, L,R,C=sinistra,destra,centro ; \leftmark = capitolo ; \rightmark = paragrafo
\fancypagestyle{plain}{%
  \fancyhead{}                                     % Pulisce il campo della testata
  \fancyfoot[CE,CO]{\thepage}       % Numero pagina in basso ed esterno in grassetto
  \fancyhead[CE,CO]{\textcolor[rgb]{0.64,0.15,0.15}{Published in \emph{Journal of the Mechanics and Physics of Solids} (2016) \textbf{91}: 204-215
\\
doi: http://dx.doi.org/10.1016/j.jmps.2016.03.011}}
\setlength{\headheight}{14.5pt}}
\newcommand{\be}[1]{\begin{equation}\label{#1}}
\newcommand{\ee}{\end{equation}}
\newcommand{\ba}[1]{\begin{eqnarray}\label{#1}}
\newcommand{\ea}{\end{eqnarray}}
\newcommand{\rf}[1]{(\ref{#1})}
\newcommand{\nn}{\nonumber}

\begin{document}

\begin{frontmatter}


\title{The destabilizing effect of external damping: \\ Singular flutter boundary for the Pfl\"uger column with vanishing external dissipation}

\author[trento]{Mirko Tommasini}
\ead{mirko.tommasini@unitn.it}
\address[trento]{Universit\'a di Trento, DICAM, via Mesiano 77, I-38123 Trento,
Italy}

\author[trento,onkmian]{Oleg N. Kirillov}
\address[onkmian]{Russian Academy of Sciences, Steklov Mathematical Institute, Gubkina st. 8,
119991 Moscow, Russia}
\ead{kirillov@mi.ras.ru}

\author[trento]{Diego Misseroni}
\ead{diego.misseroni@unitn.it}

\author[trento]{Davide Bigoni\footnote{Corresponding author: Davide Bigoni, davide.bigoni@unitn.it; +39 0461 282507}}
\ead{davide.bigoni@unitn.it}

\begin{abstract}
Elastic structures loaded by nonconservative positional forces are prone to instabilities induced by dissipation: it is well-known in fact that internal viscous damping destabilizes the marginally stable Ziegler's pendulum and Pfl\"{u}ger  column (of which  the Beck's column is a special case), two structures loaded by a tangential follower force.
The result is the so-called \lq destabilization paradox', where the critical force for flutter instability decreases by an order of magnitude  when the coefficient of internal damping becomes infinitesimally small. Until now external damping, such as that related to air drag, is believed to provide only a stabilizing effect, as one would intuitively expect. Contrary to this belief, it will be shown that the effect of external damping is qualitatively the same as the effect of internal damping, yielding a pronounced destabilization paradox. Previous results relative to destabilization by external damping of the Ziegler's and Pfl\"{u}ger's elastic structures are corrected in a definitive way leading to a new understanding of the destabilizating role played by viscous terms.

\end{abstract}

\begin{keyword}
Pfl\"uger column \sep Beck column \sep Ziegler destabilization paradox \sep external damping \sep follower force \sep mass distribution



\end{keyword}

\end{frontmatter}



%\newpage

\section{Introduction}
\label{Section1}

\subsection{A premise: the Ziegler destabilization paradox} \

In his pioneering work \cite{Ziegler1952} considered asymptotic stability of a two-linked pendulum loaded by a tangential follower force $P$, as a function of the internal damping in the viscoelastic joints connecting the two rigid and weightless bars (both of length $l$, Fig. \ref{fig1}(c)). The pendulum carries two point masses: the mass $m_1$ at the central joint and the mass $m_2$ mounted at the loaded end of the pendulum. The follower force $P$ is always aligned with the second bar of the pendulum, so that its work is non-zero along a closed path, which  provides a canonical example of a nonconservative positional force.

For two non-equal masses ($m_1=2m_2$) and null damping, Ziegler found that the pendulum is marginally stable and all the eigenvalues of the $2 \times 2$ matrix governing the dynamics are purely
imaginary and simple, if the load falls within the interval $0\leq P < P_u^-$, where
\be{ziegleru}
P_u^-=\left(\frac{7}{2}-\sqrt{2}\right) \frac{k}{l}\approx 2.086 \frac{k}{l} ,
\ee
and $k$ is the stiffness coefficient, equal for both joints.
When the load $P$ reaches the value $P_u^-$, two imaginary eigenvalues merge into a double one and the matrix governing dynamics becomes a Jordan block.
With the further increase of $P$ this double eigenvalue splits into two complex conjugate. The eigenvalue with the positive real part corresponds to a mode with an oscillating and exponentially growing amplitude, which is called flutter, or oscillatory, instability. Therefore, $P=P_u^-$ marks the onset of flutter in the {\it undamped} Ziegler's pendulum.

When the internal linear viscous damping in the joints is taken into account, Ziegler found another expression for the onset of flutter: $P=P_{i}$, where
\be{zieglerd}
P_{i}=\frac{41}{28}\frac{k}{l}+\frac{1}{2}\frac{c_i^2}{m_2l^3} ,
\ee
and $c_i$ is the damping coefficient, assumed to be equal for both joints.
The peculiarity of Eq. \rf{zieglerd} is that in the limit of vanishing damping, $c_i\longrightarrow 0$, the flutter load $P_i$ tends to the value
$41/28\,k/l \approx 1.464 \,k/l$,
considerably lower than that calculated when damping is absent from the beginning, namely, the $P_u^-$ given by Eq. (\ref{ziegleru}). This is the so-called \lq Ziegler's destabilization paradox' \citep{Ziegler1952,B1963}.

The reason for the paradox is the existence of the Whitney umbrella singularity on the boundary of the asymptotic stability domain of the dissipative system \citep{B1956,KM2007,KV2010}\footnote{
In the vicinity of this singularity, the boundary of the asymptotic stability domain is a ruled surface with a self-intersection, which corresponds to a set of marginally stable undamped systems. For a fixed damping distribution, the convergence to the vanishing damping case occurs along a ruler that meets the set of marginally stable undamped systems at a point located far from the undamped instability threshold, yielding the singular flutter onset limit for almost all damping distributions. Nevertheless, there exist particular damping distributions that, if fixed, allow for a smooth convergence to the flutter threshold of the undamped system in case of vanishing dissipation \citep{B1956,B1963,BBM1989,KV2010,K2013dg}.
}.

In structural mechanics, two types of viscous dampings are considered: (i.) one, called \lq internal', is related to the viscosity of the structural material, and (ii.) another one, called \lq external', is connected to the presence of external actions, such as air drag resistance during oscillations. These two terms enter the equations of motion of an elastic rod as proportional respectively to the fourth spatial derivative of the velocity and to the velocity of the points of the elastic line.

Of the two dissipative terms only the internal viscous damping is believed to yield the Ziegler destabilization paradox \citep{B1963,BZ1969,AY1974}.

\subsection{A new, destabilizing role for external damping} \

Differently from internal damping, {\it the role of external damping is commonly believed to be a stabilizing factor,} in an analogy with the role of stationary damping in rotor dynamics \citep{B1963,Crandall1995}. A full account of this statement together with a review of the existing results is provided in Appendix A.

Since internal and external damping are inevitably present in any experimental realization of the follower force \citep{SW1975,SKK1995,BN2011}, it becomes imperative to know how these factors affect the flutter boundary of both the Pfl\"uger column and of the Ziegler pendulum with arbitrary mass distribution.
These structures are fully analyzed in the present article, with the purpose of showing:
(i.) that external damping is a destabilizing factor, which leads to the destabilization paradox for all mass distributions;
(ii.) that surprisingly, for a finite number of particular mass distributions, the flutter loads of the externally damped structures
converge to the flutter load of the undamped case (so that only in these exceptional cases the destabilizing effect is not present); and (iii.) that the destabilization paradox is more pronounced in the case when the mass of the column or pendulum is smaller then the end mass.

Taking into account also the destabilizing role of internal damping, the  results presented in this article
demonstrate a completely new role of external damping as a destabilizing effect and
suggest that the Ziegler destabilization paradox has a much better chance of being observed in the experiments with both discrete and continuous nonconservative systems than was previously believed.

\section{Ziegler's paradox due to vanishing external damping}
\label{Section2}
\begin{figure}[!ht]
  \begin{center}
\includegraphics[width=1\textwidth]{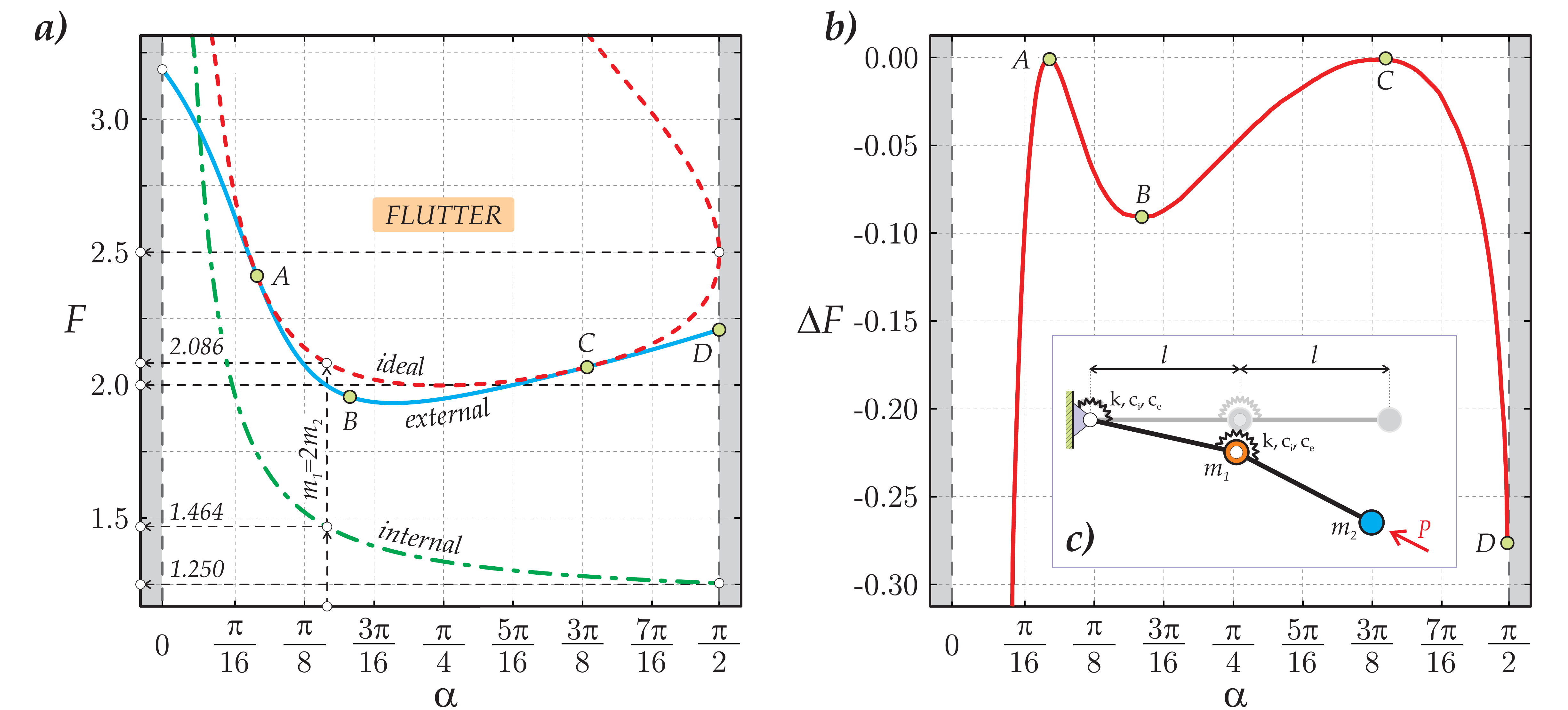}
    \caption{\footnotesize{(a) The (dimensionless) tangential force $F$, shown as a function of the (transformed via $\cot{\alpha}=m_1/m_2$) mass ratio $\alpha$, represents
				the flutter domain of (dashed/red line) the undamped, or \lq ideal', Ziegler pendulum and the flutter boundary of the dissipative system in the limit of vanishing (dot-dashed/green line) internal and (continuous/blue line) external damping. (b) Discrepancy $\Delta F$ between the critical flutter load for the ideal Ziegler pendulum and for the same structure calculated in the limit of vanishing external damping. The discrepancy quantifies the Ziegler's paradox.}}
    \label{fig1}
  \end{center}
\end{figure}

The linearized equations of motion for the Ziegler pendulum (Fig. \ref{fig1}(c)), made up of two rigid bars of length $l$, loaded by a follower force $P$, when both internal and external damping are present, have the form \citep{PI1970,Plaut1971}
\be{eqzieg}
{\bf M}\ddot {\bf x} + c_i {\bf D}_i \dot {\bf x} + c_e {\bf D}_e \dot {\bf x} +{\bf K} {\bf x} =0,
\ee
where a superscript dot denotes time derivative and $c_i$ and $c_e$ are the coefficients of internal and external damping, respectively, in front of the corresponding matrices ${\bf D}_i$ and ${\bf D}_e$
\be{matrixD}
{\bf D}_i=\left(
  \begin{array}{rr}
    2 & -1 \\
    -1 & 1 \\
  \end{array}
\right), \quad
{\bf D}_e=\frac{l^3}{6}\left(
\begin{array}{cc}
    8& 3 \\
    3 & 2\\
  \end{array}
\right),
\ee
and $\bf M$ and $\bf K$ are respectively the mass and the stiffness matrices, defined as
\be{matrixMK}
{\bf M}=\left(
    \begin{array}{rr}
      m_1l^2+m_2l^2 & m_2l^2 \\
      m_2l^2 & m_2l^2 \\
    \end{array}
  \right),\quad
{\bf K}=\left(
  \begin{array}{rr}
    -Pl+2k & Pl-k \\
    -k & k \\
  \end{array}
\right),
\ee
in which $k$ is the elastic stiffness of both viscoelastic springs acting at the hinges.

%\begin{figure}
   % \begin{center}
    %\includegraphics[angle=0, width=0.3\textwidth]{figures/ziegler_column.eps}
    %\end{center}
    %\caption{Scheme for the Ziegler pendulum.}
    %\label{ziegler_fig}
%\end{figure}

%\begin{figure}
%    \begin{center}
%    \includegraphics[angle=0, width=0.95\textwidth]{figures/eid_new.eps}
%    \end{center}
%    \caption{(a) The (dimensionless) tangential force, $F$ shown as a function of the (transformed via $\cot{\alpha}=m_1/m_2$) mass ratio $\alpha$ represents
%				the flutter domain of (red line) the undamped, or \lq ideal', Ziegler pendulum and the flutter boundary of the dissipative system in the limit of vanishing (green line) internal and (blue line) external damping. (b) Discrepancy $\Delta F$ between the critical flutter load for the ideal Ziegler pendulum and for the same structure calculated in the limit of vanishing external damping. The discrepancy quantifies the Ziegler's paradox.}
%    \label{fig1}
%    \end{figure}

%%%%%%%%%%%%%%%%%%%%%%%%%%%%%%%%%%%%%%%%%%%%%%%%%%%%%%%%%%%%%%%%%%%%%%
%%%%%%%%%%%%%%%%%%%%%%%%%%%%%%%%%%%%%%%%%%%%%%%%%%%%%%%%%%%%%%%%%%%%%%

Assuming a time-harmonic solution to the Eq. \rf{eqzieg} in the form ${\bf x}={\bf u}e^{\sigma t}$ and introducing the
non-dimensional parameters
\be{nondim}
\lambda=\frac{\sigma l}{k}{\sqrt{k m_2}},\quad E=c_e \frac{l^2}{\sqrt{km_2}}, \quad B=\frac{c_i}{l\sqrt{k m_2}}, \quad F=\frac{Pl}{k}, \quad \mu=\frac{m_2}{m_1},
\ee
an eigenvalue problem is obtained, which eigenvalues $\lambda$ are the roots of the characteristic polynomial
\ba{charpoly}
p(\lambda)&=&36\lambda^4+12(15B\mu+2E\mu+3B+E)\lambda^3+\nn\\
& &(36B^2\mu+108BE\mu+7E^2\mu-72F\mu+180\mu+36)\lambda^2+\nn\\
& &6\mu(-5EF+12B+18E)\lambda+36\mu.
\ea

In the undamped case, when $B=0$ and $E=0$, the pendulum is stable, if $0\le F < F_u^-$, unstable by flutter, if
$F_u^-\le F \le F_u^+$, and unstable by divergence, if $F>F_u^+$, where
\be{ufmu}
F_u^{\pm}(\mu)=\frac{5}{2}+\frac{1}{2\mu}\pm\frac{1}{\sqrt{\mu}}.
\ee

In order to plot the stability map for all mass distributions $0\le \mu< \infty$, a parameter
$\alpha \in [0, \pi/2]$ is introduced, so that $\cot{\alpha}=\mu^{-1}$ and hence
\be{ufa}
F_u^{\pm}(\alpha)=\frac{5}{2}+\frac{1}{2}\cot{\alpha}\pm\sqrt{\cot{\alpha}}.
\ee

The curves \rf{ufa} form the boundary of the flutter domain of the undamped, or \lq ideal', Ziegler's pendulum
shown in Fig.~\ref{fig1}(a) (red/dashed line)
in the load versus mass distribution plane \citep{Oran1972,K2011}. The smallest flutter load $F_u^-=2$ corresponds to $m_1=m_2$, i.e. to $\alpha=\pi/4$. When $\alpha$ equals $\pi/2$, the mass at the central joint vanishes ($m_1=0$) and $F_u^-=F_u^+=5/2$.
When $\alpha$ equals $\arctan{(0.5)}\approx 0.464$, the two masses are related as $m_1=2m_2$ and
$F_u^-={7}/{2}-\sqrt{2}$.

In the case when only internal damping is present ($E=0$) the Routh-Hurwitz criterion yields the flutter threshold as \citep{K2011}
\be{intf}
F_{i}(\mu,B) = \frac{25\mu^2+6\mu+1}{4\mu(5\mu+1)}+\frac{1}{2}B^2.
\ee
For $\mu=0.5$ Eq. \rf{intf} reduces to Ziegler's formula \rf{zieglerd}. The limit for vanishing internal damping is
\be{intfl}
\lim_{B \rightarrow 0}{F_{i}(\mu,B)}=F_{i}^0(\mu)=\frac{25\mu^2+6\mu+1}{4\mu(5\mu+1)}.
\ee
The limit $F_{i}^0(\mu)$ of the flutter boundary at vanishing internal damping is shown in green in Fig.~\ref{fig1}(a). Note that $F_{i}^0(0.5)=41/28$ and $F_{i}^0(\infty)=5/4$. For $0 \le \mu < \infty$ the limiting curve $F_{i}^0(\mu)$ has no common points with the flutter threshold $F_u^-(\mu)$ of the ideal system, which indicates that the internal damping causes the Ziegler destabilization paradox for \textit{every}  mass distribution.

In a route similar to the above, by employing the Routh-Hurwitz criterion, the critical flutter load of the Ziegler pendulum with the external damping $F_{e}(\mu,E)$ can be found
\ba{fe}
F_{e}(\mu,E) &=& \frac{122\mu^2-19\mu+5}{5\mu(8\mu-1)}+\frac{7(2\mu+1)}{36(8\mu-1)}E^2\nn\\
&-&(2\mu+1)\frac{\sqrt{35 E^2\mu(35E^2\mu-792\mu+360)+1296(281\mu^2-130\mu+25)}}
{180\mu(8\mu-1)}\nn
\ea
and its limit
calculated when $E \rightarrow 0$, which provides the result
\be{extfl}
F_{e}^0(\mu)=\frac{122\mu^2-19\mu+5-(2\mu+1)\sqrt{281\mu^2-130\mu+25}}{5 \mu(8\mu-1)}.
\ee
The limiting curve \rf{extfl} is shown in blue in Fig.~\ref{fig1}(a). It has a minimum $\min_{\mu}{F_{e}^0}(\mu)=-28+8\sqrt{14}\approx1.933$ at $\mu=(31+7\sqrt{14})/75\approx0.763$.

Remarkably, \textit{for almost all} mass ratios, {\it except two} (marked as A and C in Fig.~\ref{fig1}(a)),
the limit of the flutter load $F_{e}^0(\mu)$ is \textit{below} the critical flutter load $F_u^-(\mu)$ of the undamped system.
It is therefore concluded that external damping causes the discontinuous decrease in the critical flutter load exactly as it happens when internal damping vanishes. \textit{Qualitatively}, the effect of vanishing internal and external damping is \textit{the same}. The only difference is the magnitude of the discrepancy: the vanishing internal damping limit is larger than the vanishing external damping limit, see Fig.~\ref{fig1}(b), where $\Delta F(\mu)=F_{e}(\mu)-F_u^-(\mu)$ is plotted.

%%%%%%%%%%%%%%%%%%%%%%%%%%%%%%%%%%%%%%%%%%%%%%%%%%%%%%%%%%%%%%%%%%%%%%
\begin{figure}[!ht]
  \begin{center}
    \includegraphics[width=1\textwidth]{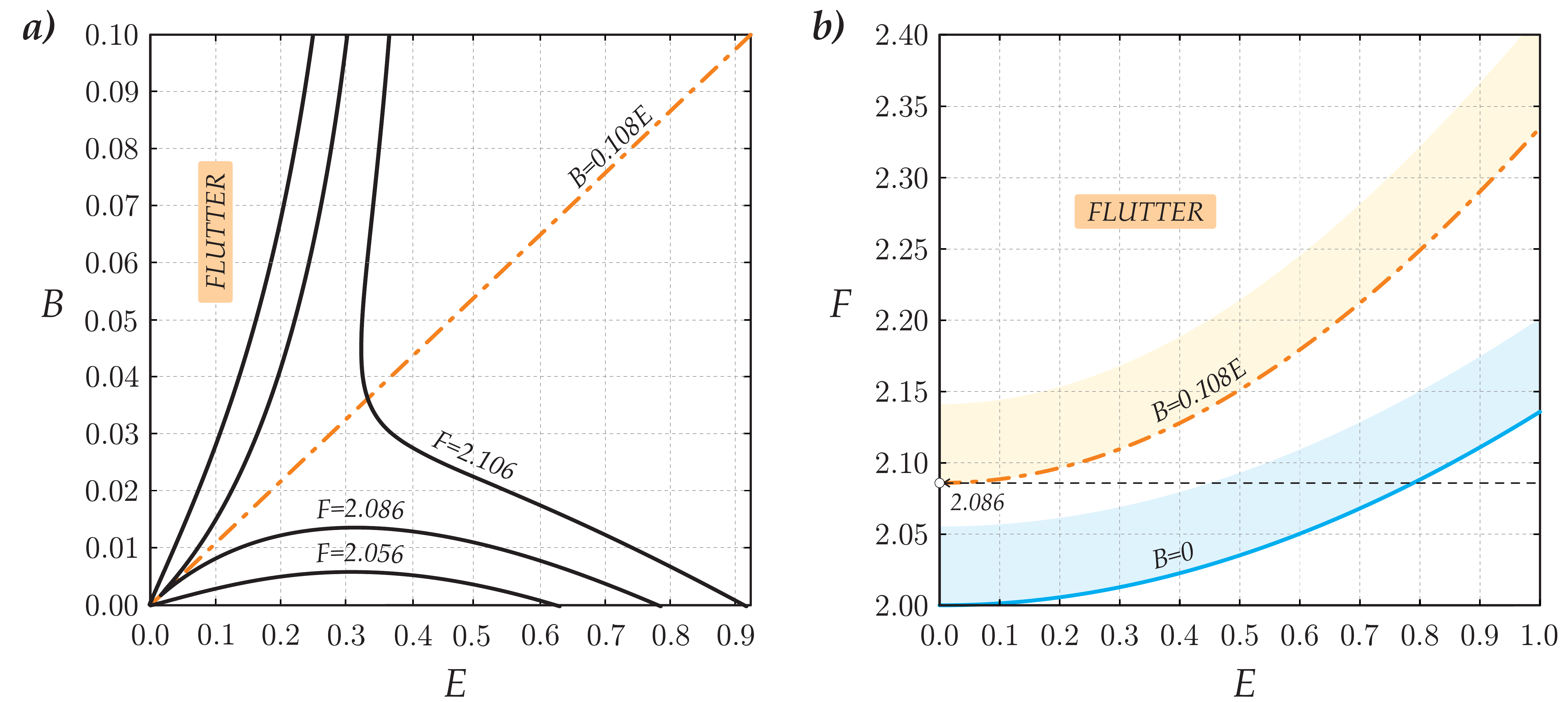}
    \caption{\footnotesize{Analysis of the Ziegler pendulum with fixed mass ratio, $\mu=m_2/m_1=1/2$: (a) contours of the flutter boundary in the internal/external damping plane, $(B,E)$, and (b) critical flutter load as a function of the external damping $E$ (continuous/blue curve) along the null internal damping line, $B=0$, and (dot-dashed/orange curve) along the line $B=\left(8/123+5\sqrt{2}/164\right)E$.}}
    \label{fig2}
  \end{center}
\end{figure}
%%%%%%%%%%%%%%%%%%%%%%%%%%%%%%%%%%%%%%%%%%%%%%%%%%%%%%%%%%%%%%%%%%%%%%

For example, $\Delta F \approx -0.091$ at the local minimum for the discrepancy, occurring at the point B with $\alpha\approx0.523$. The largest finite drop in the flutter load due to external damping occurs  at $\alpha={\pi}/{2}$, marked as point D in Fig.~\ref{fig1}(a,b):
\be{lad}
\Delta F=\frac{11}{20}-\frac{1}{20}\sqrt{281}\approx-0.288.
\ee
For comparison, at the same value of $\alpha$, the flutter load drops due to internal damping of exactly 50\%, namely, from $2.5$ to $1.25$, see Fig.~\ref{fig1}(a,b).

%%%%%%%%%%%%%%%%%%%%%%%%%%%%%%%%%%%%%%%%%%%%%%%%%%%%%%%%%%%%%%%%%%%%%%
\begin{figure}[!ht]
  \begin{center}
    \includegraphics[width=1\textwidth]{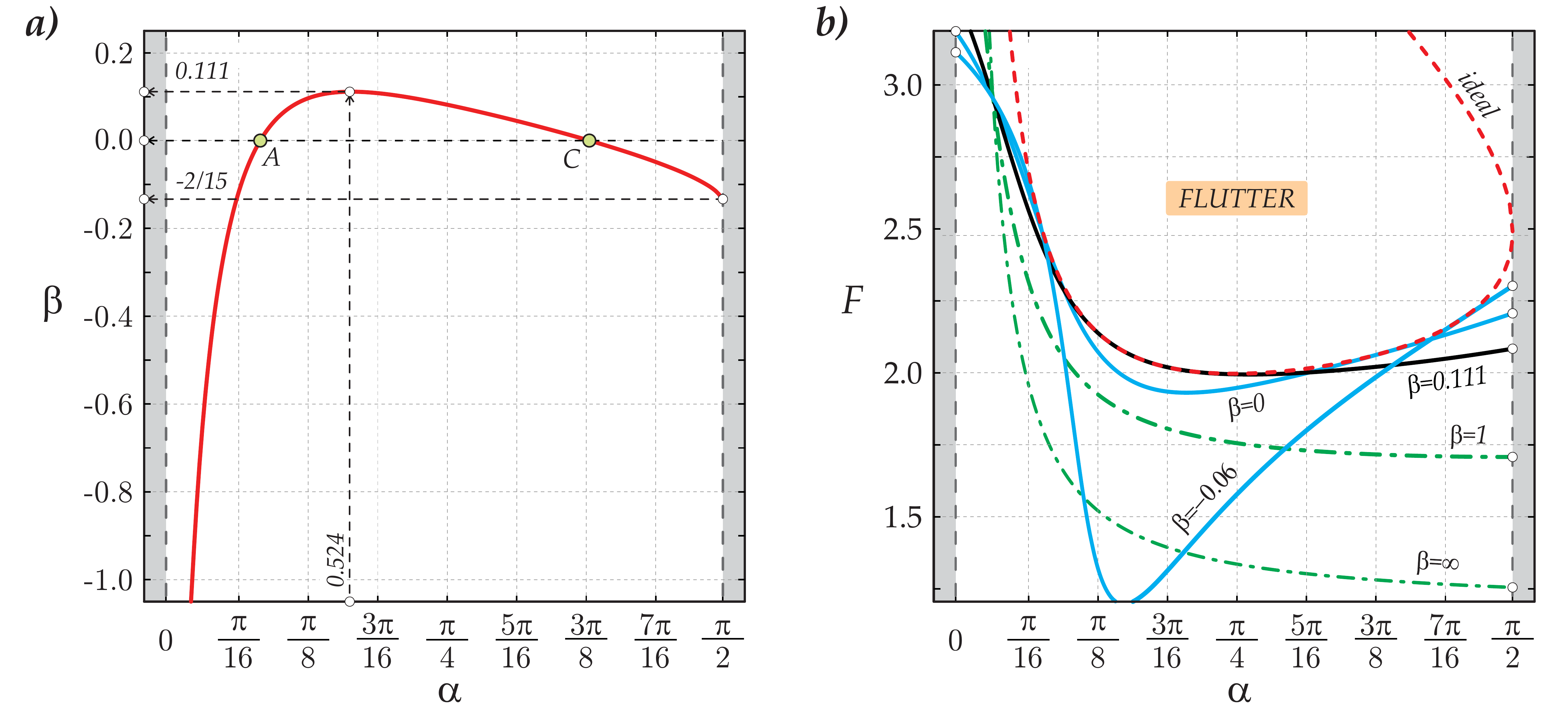}
    \caption{\footnotesize{Analysis of the Ziegler pendulum.  (a) Stabilizing damping ratios $\beta(\mu)$ according to Eq. \rf{dr} with the points A and C corresponding to the tangent points A and C in Fig.~\ref{fig1}(a) and to the points A and C of vanishing discrepancy $\Delta F=0$ in Fig.~\ref{fig1}(b). (b) The limits of the flutter boundary for different damping ratios $\beta$ have:
	  two or one or none common points with the flutter boundary (dashed/red line) of the undamped Ziegler pendulum, respectively when $\beta <0.111$ (continuous/blue curves), $\beta\approx0.111$ (continuous/black curve), and $\beta>0.111$ (dot-dashed/green curves).}}
    \label{fig3}
  \end{center}
\end{figure}
%%%%%%%%%%%%%%%%%%%%%%%%%%%%%%%%%%%%%%%%%%%%%%%%%%%%%%%%%%%%%%%%%%%%%%

As a particular case, for the mass ratio $\mu=1/2$, considered by \cite{PI1970} and \cite{Plaut1971}, the following limit flutter load is found
\be{plaut}
F_{e}^0(1/2)=2,
\ee
only slightly inferior to the value for the undamped system, $F_u^-(1/2)=7/2-\sqrt{2}\approx2.086$.
This discrepancy passed unnoticed in \citep{PI1970,Plaut1971} but gives evidence to the destabilizing effect of external damping. To appreciate this effect, the contours of the flutter boundary in the $(B,E)$ - plane are plotted in Fig.~\ref{fig2}(a) for three different values of $F$. The contours are typical of a surface with a Whitney umbrella singularity at the origin \citep{KV2010}. At $F=7/2-\sqrt{2}$ the stability domain assumes the form of a cusp with a unique tangent line, $B=\beta E$, at the origin, where
\be{cusp}
\beta=\frac{8}{123}+\frac{5}{164}\sqrt{2}\approx0.108.
\ee
For higher values of $F$ the flutter boundary is displaced from the origin, Fig.~\ref{fig2}(a), which indicates the possibility of a continuous increase in the flutter load with damping. Indeed, along the direction in the $(B,E)$ - plane with the slope \rf{cusp} the flutter load increases as
\be{tcon}
F(E)= \frac{7}{2}-\sqrt{2}+\left(\frac{47887}{242064}+\frac{1925}{40344}\sqrt{2}\right)E^2+o(E^2),
\ee
see Fig.~\ref{fig2}(b), and monotonously tends to the undamped value as $E\rightarrow 0$. On the other hand,
along the direction in the $(B,E)$ - plane specified by the equation $B=0$, the following condition is obtained
\be{bzero}
F(E)=2+\frac{14}{99}E^2+o(E^2),
\ee
see Fig.~\ref{fig2}(b), with the convergence to a lower value $F=2$ as $E\rightarrow 0$.

In general, the limit of the flutter load along the line $B=\beta E$ when $E\rightarrow 0$ is
\be{limf}
F(\beta) = \frac{504\beta^2+1467\beta+104-(4+21\beta)\sqrt{576\beta^2+1728\beta+121}}{30(1+14\beta)}\le \frac{7}{2}-\sqrt{2},
\ee
an equation showing that for almost all directions the limit is lower than the ideal flutter load. The limits only coincide in the sole direction specified by Eq. \rf{cusp}, which is different from the $E$-axis, characterized by $\beta=0$. As a conclusion, pure external damping yields the destabilization paradox even at $\mu=1/2$, which was unnoticed in \citep{PI1970,Plaut1971}.

In the limit of vanishing external ($E$) and internal ($B$) damping, a ratio of the two $\beta=B/E$ exists for which the critical load of the undamped system is attained, so that the Ziegler's paradox does not occur. This ratio can therefore be called \lq stabilizing', it exists
for every mass ratio $\mu=m_2/m_1$, and is given by the expression
\be{dr}
\beta(\mu)=-\frac{1}{3}\frac{(10\mu-1)(\mu-1)}{25\mu^2+6\mu+1}+
\frac{1}{12}\frac{(13\mu-5)(3\mu+1)}{25\mu^2+6\mu+1}\mu^{-1/2}.
\ee
Eq. (\ref{dr}) reduces for $\mu=1/2$ to Eq. \rf{cusp} and gives $\beta=-2/15$ in the limit $\mu \rightarrow \infty$.
With the damping ratio specified by Eq. \rf{dr} the critical flutter load has the following Taylor expansion near $E=0$:
\ba{femu}
F(E,\mu)&=&F_u^-(\mu)+\beta(\mu)\frac{(5\mu+1)(41\mu+7)}{6(25\mu^2+6\mu+1)}E^2\nn\\
&+&\frac{636\mu^3+385\mu^2-118\mu+25}{288(25\mu^2+6\mu+1)\mu}E^2+o(E^2),
\ea
yielding Eq. \rf{tcon} when $\mu=1/2$. Eq. (\ref{femu}) shows that the flutter load reduces to the undamped case when $E=0$ (called \lq ideal' in the figure).

%%%%%%%%%%%%%%%%%%%%%%%%%%%%%%%%%%%%%%%%%%%%%%%%%%%%%%%%%%%%%%%%%%%%%%
\begin{figure}[!ht]
  \begin{center}
    \includegraphics[width=1\textwidth]{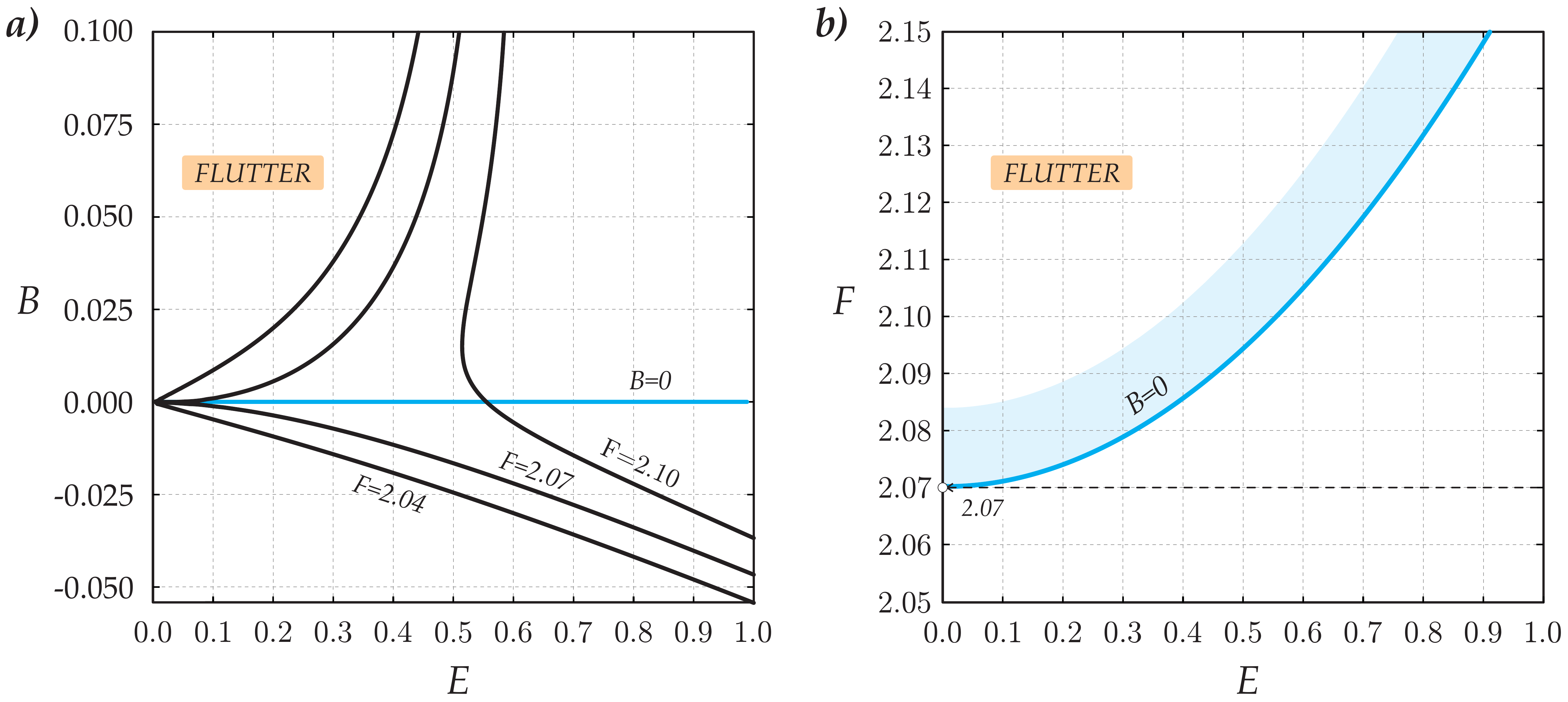}
    \caption{\footnotesize{Analysis of the Ziegler pendulum with fixed mass ratio, $\mu \approx 2.559$: (a) contours of the flutter boundary in the internal/external damping plane, $(B,E)$, and (b) critical flutter load as a function of external damping $E$ (continuous/blue curve) along the null internal damping line, $B=0$.}}
    \label{fig4}
  \end{center}
\end{figure}
%%%%%%%%%%%%%%%%%%%%%%%%%%%%%%%%%%%%%%%%%%%%%%%%%%%%%%%%%%%%%%%%%%%%%%

When the stabilizing damping ratio is null, $\beta=0$, convergence to the critical flutter load of the undamped system occurs by approaching the origin in the $(B,E)$ - plane along the $E$ - axis.
The corresponding mass ratio can be obtained finding the roots of the function $\beta(\mu)$ defined by Eq. \rf{dr}. This function has only two roots for $0\le \mu < \infty$, one at $\mu \approx0.273$ (or $\alpha \approx0.267$, marked as point A in Fig.~\ref{fig3}(a)) and another at $\mu \approx 2.559$ (or $\alpha \approx 1.198$, marked as point C in Fig.~\ref{fig3}(a)).

Therefore, if $\beta=0$ is kept in the limit when the damping tends to zero, the limit of the flutter boundary in the load versus mass ratio plane will be obtained as a curve showing two common points with the flutter boundary of the undamped system, exactly at the mass ratios corresponding to the points denoted as A and C in Fig.~\ref{fig1}(a), respectively characterized by $F\approx 2.417$ and $F \approx 2.070$.

If for instance the mass ratio at the point C is considered and  the contour plots are analyzed of the flutter boundary in the $(B,E)$ - plane, it can be noted that at the critical flutter load of the undamped system, $F\approx 2.07$, the boundary evidences a cusp with only one tangent coinciding with the $E$ axis, Fig.~\ref{fig4}(a).
It can be therefore concluded that at the mass ratio $\mu \approx 2.559$ the external damping alone has a stabilizing effect and the system does not demonstrate the Ziegler paradox due to small external damping, see Fig.~\ref{fig4}(b), where the the flutter load $F(E)$ is shown.

Looking back at the damping matrices \rf{matrixD} one may ask, what is the property of the damping operator which determines its stabilizing or destabilizing character. The answer to this question (provided by \citep{KS2005b,K2013dg} via perturbation of multiple eigenvalues) involves all the three matrices $\bf M$ (mass), $\bf D$ (damping), and $\bf K$ (stiffness). In fact, the distributions of mass, stiffness, and damping should be related in a specific manner in order that the three matrices ($\bf M$, $\bf D$, $\bf K$) have a stabilizing effect (see Appendix B for details).

\section{Ziegler's paradox for the Pfl\"uger column with external damping}
\label{Section3}

The Ziegler's pendulum is usually considered as the two-dimensional analog of the Beck column, which is a cantilevered (visco)elastic rod loaded by a tangential follower force \citep{B1952}.
Strictly speaking, this analogy is not correct because the Beck column has a different mass distribution (the usual mass distribution
of the Ziegler pendulum is $m_1=2m_2$) and this mass distribution yields different limiting behavior of the stability threshold (Section~\ref{Section2}). For this reason, in order to judge the stabilizing or destabilizing influence of external damping in the continuous case and to compare it with the case of the Ziegler pendulum, it is correct to consider the Beck column with the point mass at the loaded end, in other words the so-called \lq Pfl\"uger column' \citep{P1955}.

\begin{figure}[!ht]
  \begin{center}
    \includegraphics[width=1\textwidth]{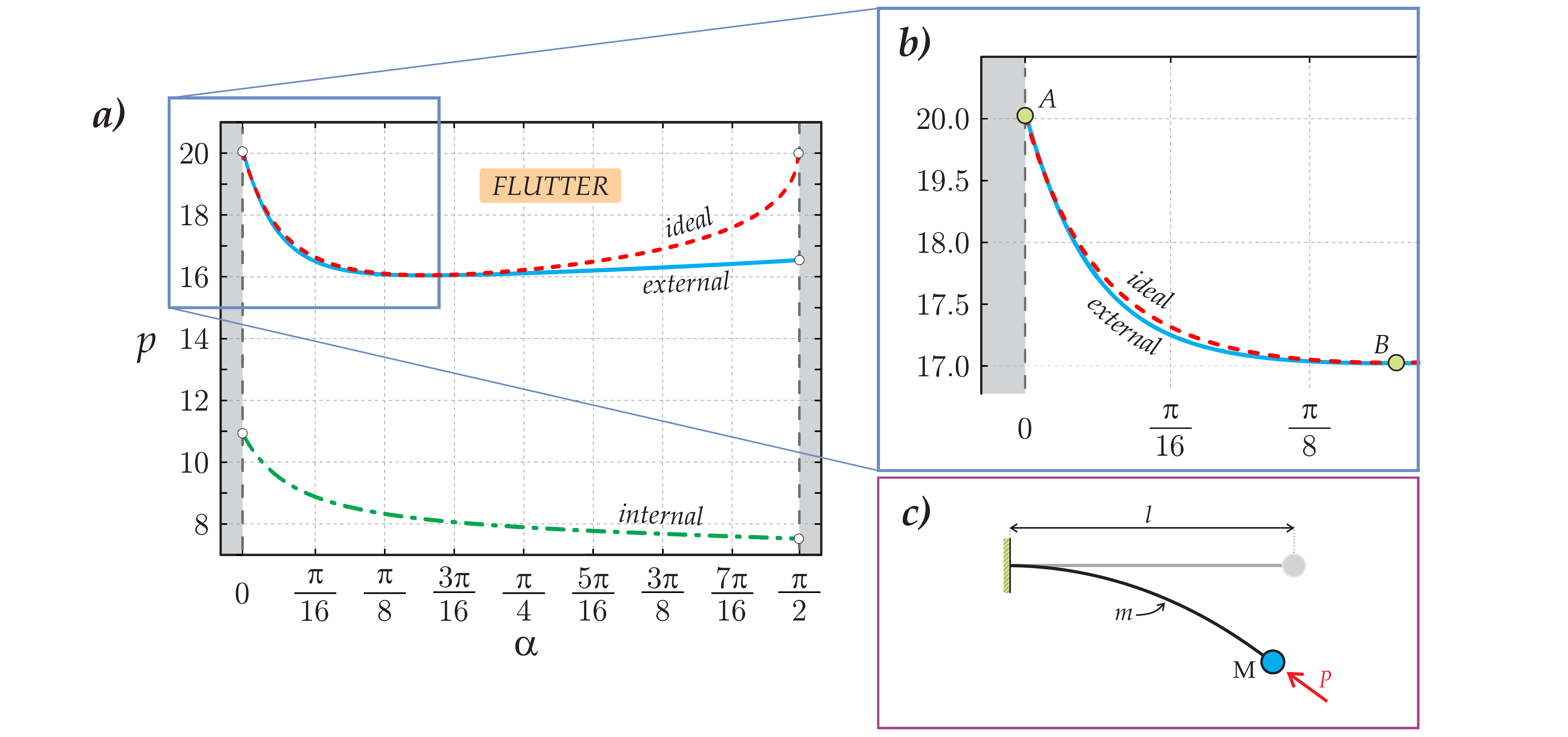}
    \caption{\footnotesize{Analysis of the
		Pfl\"uger column [scheme reported in (c)]. 		
		(a) Stability map for the Pfl\"uger's column in the load-mass ratio plane. The dashed/red curve corresponds to the stability boundary in the undamped case, the dot-dashed/green curve to the case of vanishing internal dissipation ($\gamma=10^{-10}$ and $k=0$ )  and the continuous/blue curve to the case of vanishing external damping ($k=10^{-10}$ and $\gamma=0$).
		(b) detail of the curve reported in (a) showing the destabilization effect of external damping: small, but not null.}}
    \label{fig5}
  \end{center}
\end{figure}

A viscoelastic column of length $l$, made up of a
Kelvin-Voigt material with Young modulus $E$ and viscosity modulus $E^*$,
and mass per unit length $m$ is considered, clamped
at one end and loaded by a tangential follower force $P$ at the other end (Fig.~\ref{fig5}(c)), where a point mass $M$ is mounted.

%\begin{figure}
    %\begin{center}
    %\includegraphics[angle=0, width=0.3\textwidth]{figures/pfluger_column.eps}
    %\end{center}
    %\caption{Scheme for the viscoelastic Pfl\"uger column, subject to both internal and external damping.}
    %\label{pfluger_fig}
    %\end{figure}

The moment of inertia of a cross-section of the column is denoted by $I$ and a distributed external damping is assumed, characterized by the coefficient $K$.

Small lateral vibrations of the viscoelastic Pfl\"uger column near the undeformed equilibrium state is described
by the linear partial differential equation \citep{D2003}
\be{epc}
EI \frac{\partial^4 y}{\partial x^4}+E^* I \frac{\partial^5 y}{\partial t \partial x^4}+P \frac{\partial^2 y}{\partial x^2}+K \frac{\partial y}{\partial t} + m \frac{\partial^2 y}{\partial t^2}=0,
\ee
where $y(x,t)$ is the amplitude of the vibrations and $x\in[0,l]$ is a coordinate along the column. At the clamped end $(x=0)$ Eq. \rf{epc} is equipped with the boundary conditions
\be{bcfe}
y=\frac{\partial y}{\partial x}=0,
\ee
while at the loaded end $(x=l)$, the boundary conditions are
\be{bcle}
EI\frac{\partial^2 y}{\partial x^2}+E^*I\frac{\partial^3 y}{\partial t \partial x^2}=0,\quad
EI\frac{\partial^3 y}{\partial x^3}+E^*I \frac{\partial^4 y}{\partial t \partial x^3}=M\frac{\partial^2 y}{\partial t^2}.
\ee

Introducing the dimensionless quantities
\ba{ndq}
&\xi=\frac{x}{l},\quad
\tau=\frac{t}{l^2}\sqrt{\frac{EI}{m}},\quad
p=\frac{Pl^2}{EI},\quad
\mu=\frac{M}{ml},\quad&\nn\\
&
\gamma=\frac{E^*}{El^2}\sqrt{\frac{EI}{m}},\quad
k=\frac{Kl^2}{\sqrt{mEI}}
&
\ea
and separating the time variable through $y(\xi,\tau)=l f(\xi)\exp(\lambda \tau)$, the dimensionless boundary eigenvalue problem is obtained
\ba{nde}
(1+\gamma \lambda)\partial_{\xi}^4f+p \partial_{\xi}^2 f +(k \lambda + \lambda^2)f&=&0,\nn\\
(1+\gamma \lambda)\partial_{\xi}^2f(1)&=&0,\nn\\
(1+\gamma \lambda)\partial_{\xi}^3f(1)&=&\mu \lambda^2 f(1),\nn\\
f(0)=\partial_{\xi}f(0)&=&0,
\ea
defined on the interval $\xi\in[0,1]$.

A solution to the boundary eigenvalue problem \rf{nde} was found  by \cite{Pedersen1977} and \cite{D2003} to be
\be{sol}
f(\xi)=A(\cosh(g_2 \xi)-\cos(g_1 \xi))+B(g_1\sinh(g_2 \xi)-g_2\sin(g_1 \xi))
\ee
with
\be{sol1}
g_{1,2}^2=\frac{\sqrt{p^2-4\lambda(\lambda+k)(1+\gamma\lambda)} \pm p}{2(1+\gamma \lambda)}.
\ee
Imposing the boundary conditions \rf{nde} on the solution \rf{sol} yields the characteristic equation $\Delta(\lambda)=0$ needed for the determination of the eigenvalues $\lambda$, where
\be{chare}
\Delta(\lambda)=(1+\gamma \lambda)^2A_1-(1+\gamma\lambda)A_2\mu\lambda^2
\ee
and
\ba{aa}
A_1&=&g_1g_2\left(g_1^4+g_2^4+2g_1^2g_2^2\cosh{g_2}\cos{g_1}+g_1g_2(g_1^2-g_2^2)\sinh{g_2}\sin{g_1}\right),\nn\\
A_2&=&(g_1^2+g_2^2)\left(g_1\sinh{g_2}\cos{g_1}-g_2\cosh{g_2}\sin{g_1} \right).
\ea

%%%%%%%%%%%%%%%%%%%%%%%%%%%%%%%%%%%%%%%%%%%%%%%%%%%%%%%%%%%%%%%%%%%%%%
%%%%%%%%%%%%%%%%%%%%%%%%%%%%%%%%%%%%%%%%%%%%%%%%%%%%%%%%%%%%%%%%%%%%%%

Transforming the mass ratio parameter in Eq. \rf{chare} as $\mu=\tan{\alpha}$ with $\alpha\in[0, \pi/2]$ allows the exploration of all possible ratios between the end mass and the mass of the column covering the mass ratios $\mu$ from zero $(\alpha=0)$ to infinity $(\alpha=\pi/2)$. The former case,  without end mass, corresponds to the Beck column, whereas the latter corresponds to a weightless rod with an end mass, which is known as the \lq Dzhanelidze column' \citep{B1963}.

It is well-known that the undamped Beck column loses its stability via flutter at $p\approx 20.05$ \citep{B1952}. In contrast, the undamped Dzhanelidze's column loses its stability via divergence at $p \approx 20.19$, which is the root of the equation $\tan{\sqrt{p}}=\sqrt{p}$ \citep{B1963}. These values, corresponding to two extreme situations, are connected by a marginal stability curve in the $(p,\alpha)$-plane that was numerically evaluated in \citep{P1955,B1963,Oran1972,SKK1976,Pedersen1977,RS2003}.
The instability threshold of the undamped Pfl\"uger column is shown in Fig.~\ref{fig5} as a dashed/red curve.

%%%%%%%%%%%%%%%%%%%%%%%%%%%%%%%%%%%%%%%%%%%%%%%%%%%%%%%%%%%%%%%%%%%%%%
\begin{figure}[!ht]
  \begin{center}
    \includegraphics[width=1\textwidth]{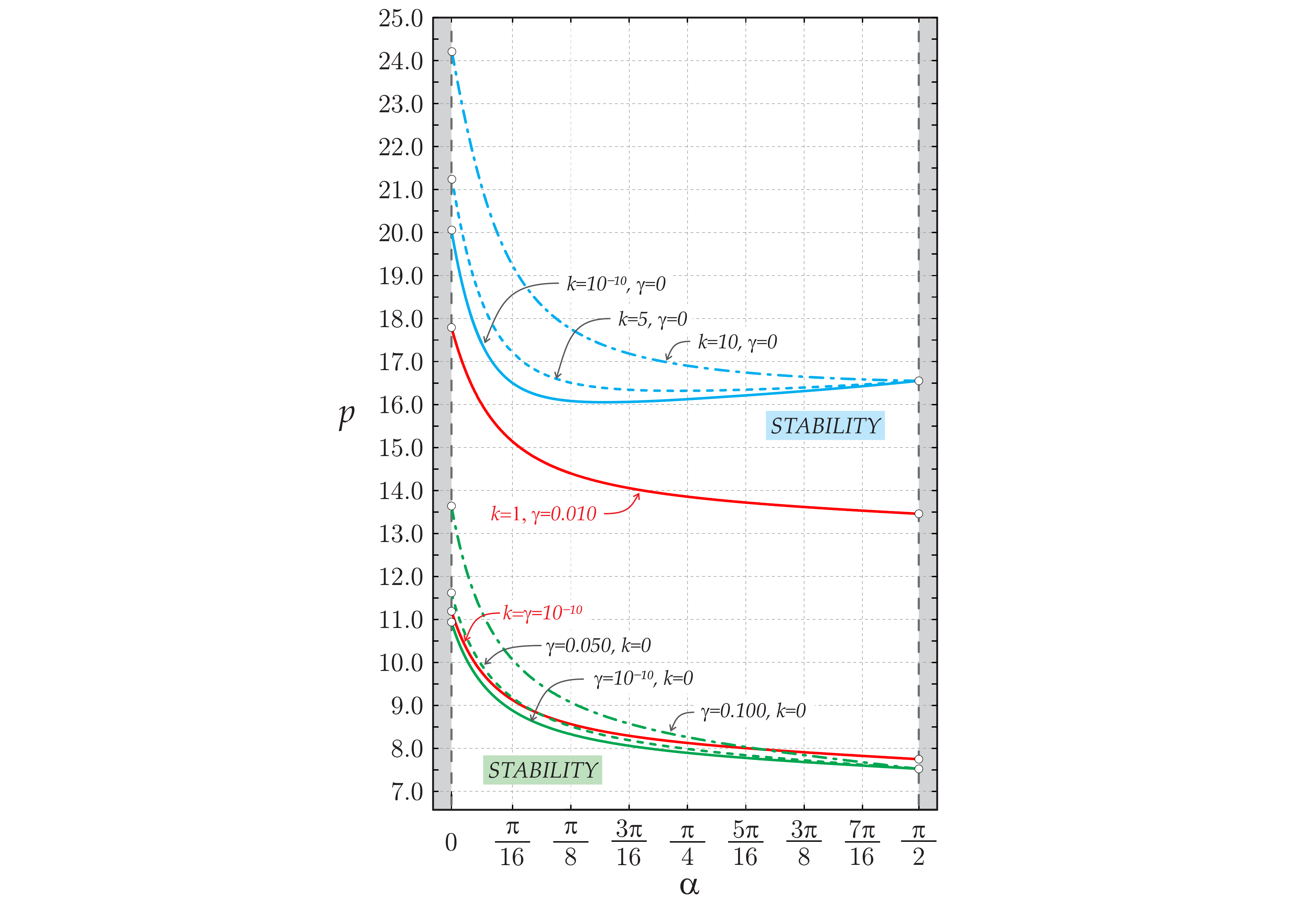}
    \caption{\footnotesize{Evolution of the marginal stability curve for the Pfl\"uger column in the $(\alpha,p)$ - plane in the case of $k=0$ and $\gamma$ tending to zero (green curves in the lower part of the graph) and in the case of $\gamma=0$ and $k$ tending to zero (blue curves in the upper part of the graph).  The cases of $k=\gamma=10^{-10}$ and of $k=1$ and $\gamma=0.01$ are
		reported with continuous/red lines.}}
    \label{fig6}
  \end{center}
\end{figure}
%%%%%%%%%%%%%%%%%%%%%%%%%%%%%%%%%%%%%%%%%%%%%%%%%%%%%%%%%%%%%%%%%%%%%%

For every fixed value $\alpha\in[0,\pi/2)$, the undamped column loses stability via flutter when an increase in $p$ causes the imaginary eigenvalues of two different modes to approach each other and merge into a double eigenvalue with one eigenfunction. When $p$ lies above the dashed/red curve, the double eigenvalue splits into two complex eigenvalues, one with the positive real part, which determines a flutter unstable mode.

At $\alpha=\pi/2$ the  stability boundary of the undamped Pfl\"uger column has a vertical tangent and the type of instability becomes divergence \citep{B1963,Oran1972,SKK1976}.

Setting $k=0$ in Eq. \rf{chare} the location in the $(\alpha, p)$-plane of the marginal stability curves can be numerically found for the viscoelastic Pfl\"uger column without external damping, but for different values of the coefficient of internal damping $\gamma$, Fig.~\ref{fig6}(a). The thresholds tend to a limit which does not share common points with the stability boundary of the ideal column, as shown in Fig.~\ref{fig5}(a), where this limit is set by the dot-dashed/green curve.

The limiting curve calculated for $\gamma=10^{-10}$ agrees well with that obtained for $\gamma=10^{-3}$ in \citep{SKK1995,RS2003}. At the point $\alpha=0$, the limit value of the critical flutter load when the internal damping is approaching zero equals the well-known value for the Beck's column, $p \approx 10.94$. At $\alpha=\pi/4$ the limiting value becomes $p\approx 7.91$, while for the case of the Dzhanelidze column $(\alpha=\pi/2)$ it becomes $p\approx7.49$.

An interesting question is
what is the limit of the stability diagram for the Pfl\"uger column in the $(\alpha,p)$-plane when the coefficient of internal damping is kept null ($\gamma=0$), while the coefficient of external damping $k$ tends to zero.

The answer to this question was previously known only for the Beck column $(\alpha=0)$, for which it was established, both numerically \citep{BZ1969,PI1970} and analytically \citep{KS2005a}, that the flutter threshold of the externally damped Beck's column is higher than that obtained for the undamped Beck's column (tending to the ideal value $p\approx 20.05$, when the external damping tends to zero). This
very particular example was at the basis of the common and incorrect opinion (maintained for decades until now) that the external damping is only a stabilizing factor, even for non-conservative loadings. Perhaps for this reason the effect of the external damping in the Pfl\"uger column has, so far, simply been ignored.

The evolution of the flutter boundary for $\gamma=0$ and $k$ tending to zero is illustrated by the blue curves in Fig.~\ref{fig6}.
It can be noted that the marginal stability boundary tends to a limiting curve which has two common tangent points with the stability boundary of the undamped Pfl\"uger column, Fig.~\ref{fig5}(b). One of the common points, at $\alpha=0$ and $p\approx 20.05$, marked as point A, corresponds to the case of the Beck column. The other corresponds to $\alpha \approx 0.516$ and $p\approx16.05$, marked as point B. Only for these two \lq exceptional' mass ratios the critical flutter load of the externally damped Pfl\"uger column coincides with the ideal value when $k\rightarrow 0$.
Remarkably, for all other mass ratios the limit of the critical flutter load for the vanishing external damping is located \textit{below} the ideal value, which means that the Pfl\"uger column fully demonstrates the \textit{Ziegler destabilization paradox due to vanishing external damping}, exactly as it does in the case of the vanishing internal damping, see Fig.~\ref{fig5}(a), where the two limiting curves are compared.

Note that the discrepancy in case of vanishing external damping is smaller than in case of vanishing internal damping, in accordance with the analogous result that was established in Section~\ref{Section2} for the Ziegler pendulum with arbitrary mass distribution. As for the discrete case, also for the Pfl\"uger column the flutter instability threshold calculated in the limit when the external damping tends to zero has only two common points with the ideal marginal stability curve. The discrepancy is the most pronounced for the case of Dzhanelidze column at $\alpha=\pi/2$, where the critical load drops from $p\approx 20.19$ in the ideal case to $p\approx 16.55$ in the case of vanishing external damping.

\section{Conclusions}
\label{Section4}

Since the finding of the Ziegler's paradox for structures loaded by nonconservative follower forces, internal damping (due to material viscosity) was considered a destabilizing factor, while external damping (due for instance to air drag resistance) was believed to merely provide a stabilization. This belief originates from results obtained only for the case of Beck's column, which does not carry an end mass.
This mass is present in the case of the Pfl\"uger's column, which was never analyzed before from the point of view of the Ziegler paradox.
A revisitation of the Ziegler's pendulum and the analysis of the Pfl\"uger column has revealed that the Ziegler destabilization paradox occurs as related to the vanishing of the external damping, no matter what is the ratio between the end mass and the mass of the structure. Results presented in this article clearly show that the destabilizing role of external damping was until now misunderstood, and that experimental proof of the destabilization paradox in a mechanical laboratory is now more plausible than previously thought. Moreover, the fact that external damping plays a destabilizing role may have important consequences in structural design and this opens new perspectives for energy harvesting devices.

\section*{Acknowledgements}
The authors gratefully acknowledge financial support from the ERC Advanced Grant ‘Instabilities and nonlocal multiscale modelling of materials’ FP7-PEOPLE-IDEAS-ERC-2013-AdG (2014-2019).





%%%%%%%%%%%%%%%%%%%%%%%%%%%%%%%%%%%%%%%%%%%%%%%%%%%%%%%%%%%%%%%%%%%%

\clearpage
\setcounter{equation}{0}
\renewcommand{\theequation}{{A}.\arabic{equation}}
\begin{center}

{\bf Appendix A. - The stabilizing role of external damping and the destabilizing role of internal damping}\\

\end{center}

\noindent
A critical review of the relevant literature is given in this Appendix, with the purpose of explaining the historical origin of the misconception that the external damping introduces a mere stabilizing effect for structures subject to flutter instability.

\cite{PI1970} considered the Ziegler pendulum with $m_1=2m_2$, without internal damping (in the joints), but subjected to an external damping proportional to the velocity along the rigid rods of the double pendulum\footnote{
Note that different mass distributions were never analyzed in view of external damping effect. In the absence of damping, stability investigations were carried out by  \cite{Oran1972} and \cite{K2011}.
}. In this system the critical flutter load increases with an increase in the external damping, so that they presented a plot showing that the flutter load converges to a value which is very close to $P_u^-$. However, they did not calculate the critical value in the {\it limit} of vanishing external damping, which would have revealed a value slightly smaller than the value corresponding to the undamped system\footnote{
In fact, the flutter load of the externally damped Ziegler pendulum with $m_1=2m_2$, considered by \cite{PI1970} and \cite{Plaut1971} tends to the value $P=2$ which is smaller than $P_u^-\approx 2.086$, therefore revealing the paradox.
}.
In a subsequent work, \cite{Plaut1971} confirmed his previous result and demonstrated that internal damping with equal damping coefficients destabilizes the Ziegler pendulum,  whereas external damping has a stabilizing effect, so that it does not lead to the destabilization paradox. \cite{Plaut1971} reports a stability diagram (in the external versus internal damping plane) that implicitly indicates the existence of the Whitney umbrella singularity on the boundary of the asymptotic stability domain. These conclusions agreed with
other studies on the viscoelastic cantilevered Beck's column \citep{B1952}, loaded by a follower force which displays the paradox only for internal Kelvin-Voigt damping \citep{BZ1969,PI1970,AY1974, KS2005a} and were supported by studies on the abstract settings \citep{Done1973,Walker1973,KS2005b}, which have proven the stabilizing character of external damping, assumed to be proportional to the mass
\citep{B1963,Zhinzher1994}.

The Pfl\"uger column [a generalization of the Beck problem in which a concentrated mass is added to the loaded end, \cite{P1955}, see also \cite{SKK1976}, \cite{Pedersen1977}, and \cite{Chen1992}] was analyzed by \cite{SKK1995} and \cite{RS2003}, who numerically found that the internal damping leads to the destabilization paradox
for all ratios of the end mass to the mass of the column. The role of external damping  was investigated only by \cite{D2003} who concludes that large external damping provides a stabilizing effect.

The stabilizing role of external damping was questioned only in the work by \cite{PS1987}, in which the Ziegler pendulum and the Beck column were considered  with a dash-pot damper attached to the loaded end (a setting in which the external damper can be seen as something different than an air drag, but as merely an additional structural element, as suggested by \cite{Zhinzher1994}).
In fact the dash-pot was shown to always yield the destabilization paradox, even in the presence of internal damping, no matter what the ratio is between the coefficients of internal and external damping \citep{KS2005c,K2013dg}.

In summary, there is a well-established opinion that external damping stabilizes structures loaded by nonconservative positional forces.

\begin{center}

{\bf Appendix B. - A necessary condition for stabilization of a general 2 d.o.f. system}\\
\end{center}

\noindent

\cite{KS2005b} considered the stability of the system
\be{b1}
{\bf M}\ddot{\bf x}+\varepsilon {\bf D} \dot{\bf x} + {\bf K}{\bf x}=0,
\ee
where $\varepsilon >0$ is a small parameter and ${\bf M}={\bf M}^T$, ${\bf D}={\bf D}^T$, and ${\bf K}\ne{\bf K}^T$ are real matrices of order $n$. In the case $n=2$, the characteristic polynomial of the system \rf{b1},
$$
q(\sigma,\varepsilon)=\det({\bf M}\sigma^2+\varepsilon {\bf D} \sigma + {\bf K}),
$$
can be written by means of the Leverrier algorithm (adopted for matrix polynomials by \cite{WL1993}) in a compact form:
\be{b2}
q(\sigma,\varepsilon)=\det {\bf M} \sigma^4+\varepsilon{\rm tr}({\bf D}^* {\bf M})\sigma^3+({\rm tr}({\bf K}^* {\bf M})+\varepsilon^2\det {\bf D})\sigma^2+\varepsilon{\rm tr}({\bf K}^* {\bf D}) \sigma +\det {\bf K},
\ee
where ${\bf D}^*={\bf D}^{-1} \det{\bf D}$ and ${\bf K}^*={\bf K}^{-1} \det{\bf K}$ are adjugate matrices and ${\rm tr}$ denotes the trace operator.

Let us assume that at $\varepsilon = 0$ the undamped system \rf{b1} with $n=2$ degrees of freedom be on the flutter boundary, so that its eigenvalues are imaginary and form a double complex-conjugate pair $\sigma=\pm i\omega_0$ of a Jordan block. In these conditions, the real critical frequency $\omega_0$ at the onset of flutter follows from $q(\sigma,0)$ in the closed form \citep{K2013dg}
\be{b3}
\omega_0^2=\sqrt{\frac{\det{\bf K}}{\det{\bf M}}}.
\ee

A dissipative perturbation $\varepsilon {\bf D}$ causes splitting of the double eigenvalue $i\omega_0$, which is described by the Newton-Puiseux series  $\sigma(\varepsilon)=i\omega_0\pm i \sqrt{h \varepsilon}+o(\varepsilon)$,
where the coefficient $h$ is determined in terms of the derivatives of the polynomial $q(\sigma,\varepsilon)$ as
\be{b4}
h:=\left.\frac{d q }{d \varepsilon}\left(\frac{1}{2}\frac{\partial^2 q}{\partial \sigma^2} \right)^{-1}\right |_{\varepsilon=0,\,\sigma=i\omega_0}=\frac{{\rm tr}({\bf K}^* {\bf D})-\omega_0^2{\rm tr}({\bf D}^* {\bf M}) }{4i\omega_0\det {\bf M} }.
\ee
Since the coefficient $h$ is imaginary, the double eigenvalue $i\omega_0$ splits generically into two complex eigenvalues, one of them with the positive real part yielding flutter instability \citep{KS2005b}. Consequently, $h=0$ represents a \textit{necessary condition} for $\varepsilon {\bf D}$ to be a \textit{stabilizing perturbation} \citep{KS2005b}.

In the case of the system \rf{eqzieg}, with matrices \rf{matrixMK}, it is readily obtained
\be{b5}
\omega_0^2=\frac{k}{l^2\sqrt{m_1m_2}}.
\ee

Assuming ${\bf D}={\bf D}_i$, eq. \rf{b4} and the representations \rf{matrixMK} and \rf{b5} yield
\be{b6}
h=h_i:=\frac{i}{m_1 l^2}\frac{5\mu-2\sqrt{\mu}+1}{4\mu },
\ee
so that the equation $h_i=0$ has as solution the complex-conjugate pair $\mu=(-3\pm4i)/25$. Therefore, for \textit{every} real mass distribution $\mu \ge 0$ the dissipative perturbation with the matrix ${\bf D}={\bf D}_i$ of internal damping results to be destabilizing.

Similarly, eq. \rf{b4} with ${\bf D}={\bf D}_e$ and representations \rf{b5}, \rf{matrixMK}, and $F=F_u^-(\mu)$ yield
\be{b6}
h=h_e:=\frac{il}{48m_1}\frac{8\mu^2-11\sqrt{\mu^3}-6\mu+5\sqrt{\mu}}{\mu^2},
\ee
so that the constraint $h_e=0$ is satisfied only by the two following real values of $\mu$
\be{b7}
\mu_A\approx0.273,\quad \mu_C\approx 2.559.
\ee
The mass distributions \rf{b7} correspond exactly to the points A and C in Fig.~\ref{fig1}, which are common
for the flutter boundary of the undamped system and for that of the dissipative system in the limit of vanishing external damping. Consequently, the dissipative perturbation with the matrix ${\bf D}={\bf D}_e$ of external damping can have a stabilizing effect for \textit{only two} particular mass distributions \rf{b7}. Indeed, as it is shown in the present article, the external damping is destabilizing for every $\mu\ge0$, except for $\mu=\mu_A$ and $\mu=\mu_C$.

Consequently, the stabilizing or destabilizing effect of damping with the given matrix $\bf D$ is determined not only by its spectral properties, but also by how it `interacts' with the mass and stiffness distributions. The condition which selects possibly stabilizing triples ($\bf M$, $\bf D$, $\bf K$) in the general case of $n=2$ degrees of freedom is therefore the following
\be{b8}
{\rm tr}({\bf K}^* {\bf D})=\omega_0^2{\rm tr}({\bf D}^* {\bf M}).
\ee

\end{document}